\documentclass
[rmp,twocolumn,amsmath,letterpaper,raggedbottom,nobalancelastpage]{revtex4}

\usepackage[german,english]{babel}
\usepackage{ifpdf}
\ifpdf
  \def\hrefx#1#2#3{\href{#1}{{#2}\ {#3}}}
\else
 \def\hrefx#1#2#3{\href{#1}{#2}\penalty0\ \href{#1}{#3}}
\fi

\usepackage{times}
\usepackage[bookmarks=false]{hyperref}
\usepackage{array}

\hypersetup{pdftitle={The calculation of longitude and latitude
from geodesic measurements},
pdfauthor={F. W. Bessel},
pdfsubject={Astron. Nachr. 4(86), 241-254 (1825)},
pdfkeywords={doi:10.1002/asna.18260041601}}

\usepackage{graphicx}

\makeatletter\def\raggedcolumn@skip{\vskip\z@\@plus.0001fil\relax}\makeatother

\newcommand{\p}{\!\cdot\!}
\def\dms#1:#2:#3:{#1^\circ\,#2'\,#3''}
\newcommand{\colog}{\mathop{\mathrm{colog}}\nolimits}
\def\sumbar#1{\overline{\vphantom{\rule{0pt}{2.5ex}}#1}}
\let\b=\bar
\def\d{d}

\date[Originally published: ]{October 1825;
translated: August 13, 2009; revised: March 29, 2012}

\begin{document}

\title{The calculation of longitude and latitude from geodesic
measurements\footnote
{This is an English translation of \hrefx
{http://adsabs.harvard.edu/full/1825AN......4..241B}
{\selectlanguage{german}\it "Uber die Berechnung der geographischen}
{\selectlanguage{german}\it
L"angen und Breiten aus geod"atischen Vermessungen},
{\selectlanguage{german}Astronomische Nachrichten}
{\bf 4}(86), 241--254 (1825),
doi:\href{http://dx.doi.org/10.1002/asna.18260041601}
{10.1002/asna.18260041601}.
The paper also appears in
\href{http://books.google.com/books?id=vX4EAAAAYAAJ&pg=PA5}
{{\selectlanguage{german}\it
Abhandlungen von Friedrich Wilhelm Bessel}, Vol.~3},
pp.~5--14 (W. Engelmann, Leipzig, 1876).
The translation has been prepared and edited by
Charles F. F. Karney
$\langle$\href{mailto:charles.karney@sri.com}{charles.karney@sri.com}$\rangle$
and Rodney E. Deakin
$\langle$\href{mailto:rod.deakin@rmit.edu.au}{rod.deakin@rmit.edu.au}$\rangle$,
with the assistance of Max Hunter and Stephan Brunner.  The mathematical
notation has been updated to conform to current conventions and, in a
few places, the equations have been rearranged for clarity.  Several
errors have been corrected, a figure has been included, and the tables
have been recomputed.  A transcription of the original paper with the
updated mathematical notation and with the corrections is available at
\href{http://arxiv.org/abs/0908.1823}{arXiv:0908.1823}.
A contemporary, but partial, translation into English appeared in
Quart.\ Jour.\ Roy.\ Inst.\ {\bf 21}(41), 138--152 (1826).}}

\author{F. W. Bessel}
\affiliation{K\"onigsberg Observatory}

\maketitle

\catcode`\@=\active
\def @{.}

\section{Introduction} \label{sec1}

Consider a geodesic line between two points $A$ and $B$ on the surface
of the Earth.  Given the position of $A$, the length of the line and its
azimuth at $A$, we wish to determine the position of $B$ and the azimuth
of the line there.  This problem occurs so frequently that I undertook
to construct tables to simplify the computation.  In order to explain
the method clearly, I start by deriving the fundamental properties of
geodesic lines on a spheroid of revolution.  Even though aspects of this
derivation may already be well known, the benefit of having the entire
development presented together outweighs the cost of repeating
it.\footnote
{In Secs.~\ref{sec2}--\ref{sec4}, Bessel gives a concise summary of the
work of several other authors, notably, Clairaut, Euler, du S\'ejour, Legendre,
and Oriani.  Bessel's contributions, which start in Sec.~\ref{sec5},
consist of his methods for expanding the distance and longitude
integrals and his compilation of tables to provide a practical method
for computing geodesics.  Two sentences have been omitted from this
translation of the introduction.  In one, Bessel refers to two letters
he published earlier in the {\selectlanguage{german}\it Astronomische
Nachrichten} which do not, however, have a direct bearing on the
present work.  In the other, he criticizes ``du S\'ejour's method,'' but
without providing details; in any case, such criticism is misplaced
because du S\'ejour had died over 30 years earlier and Bessel does not
cite more recent work.}

\section{The characteristic equation for a geodesic} \label{sec2}

Take two points $A$ and $B$ on the surface on a spheroid\footnote
{``Spheroid'' here is used in the sense of a shape approximating a
sphere.  Sections \ref{sec2} and \ref{sec3} treat the case of a
rotationally symmetric earth.  In Sec.~\ref{sec4}, Bessel specializes to
a rotationally symmetric ellipsoid.}
of revolution joined by some specified curve.  Consider two neighboring
points on the curve with latitudes $\phi$ and $\phi+\d\phi$ and
longitudes relative to $A$ of $w$ and $w + \d w$ (measuring east
positive).  Let the distance between them be $\d s$, the azimuth of the line
directed toward $A$ be $\alpha$ (measured clockwise from north), the
radius of the circle of latitude be $r$, and the meridional radius of
curvature be $R$; then we find\footnote
{The minus signs appear in (\ref{eq1}) because $\alpha$ is the back
azimuth, pointing to $A$, while $\d s$ advances the geodesic away from
$A$.  In this section, Bessel assumes an easterly geodesic so that
$\d s/\d w > 0$.  However the final result, Eq.~(\ref{eq2}), is general.}
\begin{equation}\label{eq1}
\begin{split}
\cos\alpha\,\d s & = -R\,\d\phi = \frac{\d r}{\sin\phi},\\
\sin\alpha\,\d s & = -r\,\d w,
\end{split}
\end{equation}
which gives
\[
\d s = \sqrt{R^2\,\d\phi^2+r^2\,\d w^2}.
\]
If we write $p$ for $\d\phi/\d w$ and $U$ for $\sqrt{R^2p^2+r^2}$, this
becomes
\[
\d s = U\,\d w.
\]
The distance along the curve between the two points $A$ and $B$ is
therefore
\[
s = \int U\,\d w,
\]
where the integration is from $A$ to $B$.  If the curve is the geodesic
or {\it shortest} path, then the relation between $\phi$ and $w$ must be
such that the integral is a minimum.  If we perturb this relation so
that $\phi$ is replaced by $\phi + z$ where $z$ is an arbitrary function
of $w$ which vanishes at the end points (because these points lie on
both curves), then the perturbed length,
\[
s' = \int U'\,\d w,
\]
must be larger than $s$ for all $z$.

Expanding $U(\phi,p)$ in a Taylor series, we obtain\footnote
{The notation here employs partial derivatives instead of Bessel's less
formal use of differentials.}
\[
U' = U + \frac{\partial U}{\partial\phi}z +
\frac{\partial U}{\partial p}\frac{\d z}{\d w} + \ldots
\]
and therefore we have
\[
s' = s + \int\biggl(
\frac{\partial U}{\partial\phi}z +
\frac{\partial U}{\partial p}\frac{\d z}{\d w}\biggr)\,\d w
+ \ldots,
\]
where we have explicitly included terms only up to first order in $z$.
For $s$ to be a minimum, we require that
\[
\int\biggl(
\frac{\partial U}{\partial\phi}z +
\frac{\partial U}{\partial p}\frac{\d z}{\d w}\biggr)\,\d w
+ \ldots\ge0
\]
for all $z$.  Since this must also hold if $z$ is replaced by $-z$ and
since we can take $z$ so small that the first order terms are bigger
that the sum of all the higher order terms (except if the first order
terms vanish), it follows that the condition that $s$ be minimum is
\[
\int\biggl(
\frac{\partial U}{\partial\phi}z +
\frac{\partial U}{\partial p}\frac{\d z}{\d w}\biggr)\,\d w=0.
\]
Integrating the second term by parts to give
$z({\partial U}/{\partial p}) -
\int z[\d({\partial U}/{\partial p})/\d w]\,\d w$
and remembering that $z$ vanishes at the end points, we obtain
\[
\int z\biggl\{
\frac{\partial U}{\partial\phi} -
\frac\d{\d w}\biggl(\frac{\partial U}{\partial p}\biggr)\biggr\}\,\d w
=0.
\]
Since this integral must vanish for {\it arbitrary} $z$, we
find\footnote
{This is the Euler-Lagrange equation of the calculus of variations.}
\[
\frac{\partial U}{\partial\phi} -
\frac\d{\d w}\biggl(\frac{\partial U}{\partial p}\biggr)
=0
\]
or, multiplying by $\d\phi/\d w = p$,
\[
\frac{\partial U}{\partial\phi}\frac{\d\phi}{\d w}
+\frac{\partial U}{\partial p}\frac{\d p}{\d w}
-\frac{\d p}{\d w}\frac{\partial U}{\partial p}
- p\,\frac\d{\d w}\biggl(\frac{\partial U}{\partial p}\biggr)=0,
\]
which on integrating with respect to $w$ becomes
\[
 U - p\biggl(\frac{\partial U}{\partial p}\biggr)
= \mathrm{const.}
\]
Substituting $\sqrt{r^2+R^2p^2}$ for $U$, we obtain\footnote
{A. C. Clairaut gives a geometric derivation of this result in
\hrefx{http://books.google.com/books?id=GOAEAAAAQAAJ&pg=PA406}
{M\'em.\ de}{l'Acad.\ Roy.\ des Sciences de Paris 1733, 406--416 (1735)}.
The equation also follows from conservation of angular momentum for a
mass sliding without friction on a spheroid of revolution.}
\[
 \frac r{\sqrt{1+(R^2/r^2)p^2}} =
-r\sin\alpha
=\mathrm{const.},
\]
which is the well known characteristic equation of the geo\-desic.

If the azimuth of the geodesic at $A$ (in the direction of $B$) is
$\alpha'$ and the distance of $A$ from the rotation axis is $r'$, we
have
\[
r'\sin(\alpha'+180^\circ) = r\sin\alpha,
\]
or
\begin{equation}\label{eq2}
r'\sin\alpha' = -r\sin\alpha.
\end{equation}

\section{The auxiliary sphere} \label{sec3}

Let the maximum distance of the spheroid to the rotation axis be $a$, so
that $r$ and $r'$ are less than or equal to $a$; we can then
write\footnote
{The quantity $u$ is the {\it reduced} or {\it parametric} latitude.}
\[
r'= a\cos u',\quad r = a\cos u,
\]
and equation (\ref{eq2}) becomes
\begin{equation}\label{eq3}
\cos u' \sin\alpha'= - \cos u \sin\alpha.
\end{equation}
\begin{figure}[t]
\begin{center}
\ifpdf
    \vspace{-5mm}\mbox{\hspace{-13mm}\includegraphics
      [width=94mm,angle=0]{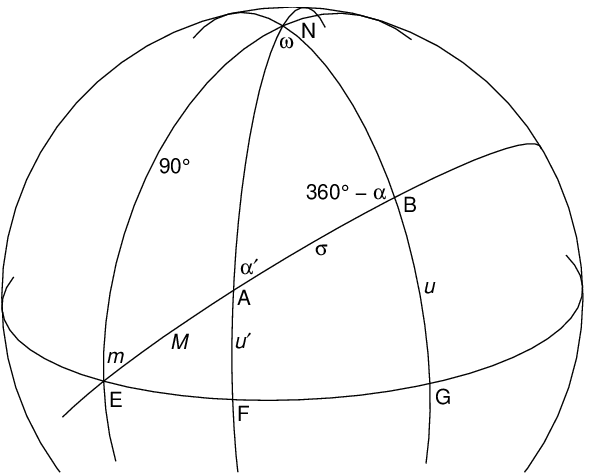}\hspace{-10mm}}\vspace{-11mm}
\else
    \vspace{1mm}\mbox{\hspace{0mm}\includegraphics
      [width=74mm,angle=0]{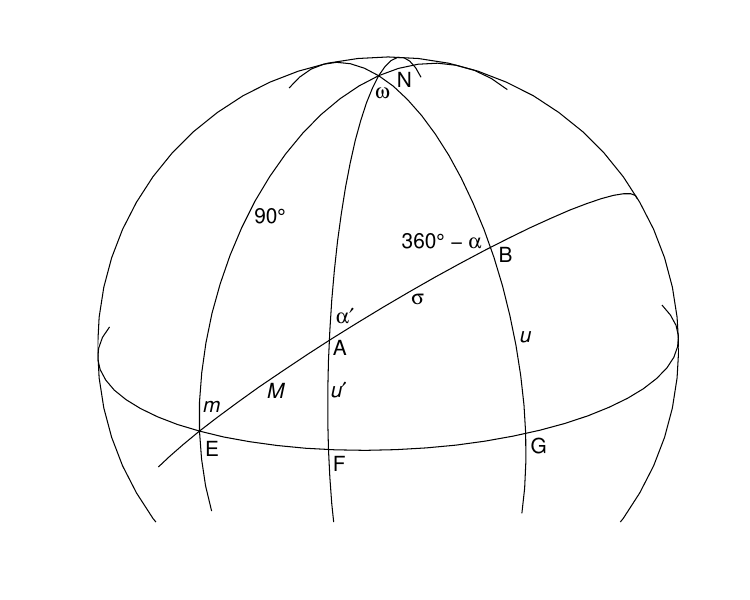}}\vspace{-1mm}
\fi
\end{center}
\caption
{Spherical triangles on the auxiliary sphere.  $EAB$ is the geodesic,
$N$ is the pole; $EFG$ is the equator; and $N\!E$, $N\!AF$, and $N\!BG$
are meridians.}
\label{figtri}
\end{figure}%
This equation relates two sides of a spherical triangle,\footnote
{See the triangle $ABN$ on the ``auxiliary sphere'' in Fig.~\ref{figtri};
Equation (\ref{eq3}) is the sine rule applied to angles $A$ and $B$ of
the triangle.}
$90^\circ - u'$ and $90^\circ - u$, and their opposite angles,
$360^\circ - \alpha$ and $\alpha'$.  The third side $\sigma$ and its
opposite angle $\omega$ will appear in the following calculations giving
elegant expressions for the joint variations of $s$, $u$ and $w$.  In
particular, using the well known differential formulas of spherical
trigonometry, we find\footnote
{Here and in the rest of the paper, the differentials give the movement
of point $B$ along the geodesic defined with point $A$ and $\alpha'$
held fixed.}
\begin{align*}
\d u &= -\cos\alpha\,\d\sigma,\\
\cos u\,\d\omega &= -\sin\alpha\,\d\sigma.
\end{align*}
Substituting these in equations (\ref{eq1}) and expressing $r$ in terms
of $u$ gives
\begin{equation}\label{eq4}
\begin{split}
\d s &=a \frac{\sin u}{\sin \phi} \d\sigma,\\
\d w &= \frac{\sin u}{\sin \phi} \d\omega.
\end{split}
\end{equation}

\section{The equations for a geodesic on an ellipsoid} \label{sec4}

I now assume that the meridian is an ellipse with equatorial radius
$a$, polar semi-axis $b$, and eccentricity $e=\sqrt{a^2-b^2}/a$.\footnote
{In Bessel's time, it was known that the earth could be approximated by
an oblate ellipsoid, $a > b$, but the eccentricity had not been
determined accurately.  Therefore, Bessel computes tables which are
applicable to ellipsoids with a range of eccentricities.}
The equation for an ellipse expressed in terms of cartesian coordinates
is
\[
 \frac{x^2}{a^2} + \frac{y^2}{b^2}
= 1.
\]
Differentiating this and setting $\d y/\d x = -\cot\phi$, we obtain
\[
 \frac{x\sin\phi}{a^2} - \frac{y\cos\phi}{b^2}
= 0;
\]
eliminating $y$ between these equations then gives
\[
x = \frac{a\cos\phi}{\sqrt{1-e^2\sin^2\phi}}.
\]
The quantity $x$ is the same as $r = a \cos u$, which gives the
relationships between $\phi$ and $u$,
\begin{align*}
\cos u &= \frac{\cos\phi}{\sqrt{1-e^2\sin^2\phi}},&
\cos\phi &= \frac{\cos u\sqrt{1-e^2}}{\sqrt{1-e^2\cos^2u}},\\
\sin u &= \frac{\sin\phi\sqrt{1-e^2}}{\sqrt{1-e^2\sin^2\phi}},&
\sin\phi &= \frac{\sin u}{\sqrt{1-e^2\cos^2u}},\\
\tan u &= \tan\phi\sqrt{1-e^2},&
\tan\phi &= \frac{\tan u}{\sqrt{1-e^2}},
\end{align*}
and
\[
\frac{\sin u}{\sin\phi} = \sqrt{1-e^2\cos^2u}.
\]
Substituting this into (\ref{eq4}), we obtain the differential equations
for a geodesic on an ellipsoid
\begin{equation}\label{eq5}
\begin{split}
\d s & = a \sqrt{1-e^2\cos^2u}\,\d\sigma,\\
\d w & = \sqrt{1-e^2\cos^2u}\,\d\omega.
\end{split}
\end{equation}

\section{The distance integral} \label{sec5}

To integrate the first of these differential equations, I use the three
relations between $u'$, $u$, $\alpha'$, $\alpha$ and $\sigma$,\footnote
{Referring to Fig.~\ref{figtri}, consider two central cartesian
coordinate systems with the $xy$ plane containing the geodesic $EAB$,
and either $A$ or $B$ lying on the $x$ axis.  Equations (\ref{eq6}) give
the transformation between the coordinates of $N$ in the two systems,
$[\sin u', \cos u'\cos\alpha', \cos u'\sin\alpha']$ and $[\sin u, -\cos
u\cos\alpha, -\cos u\sin\alpha]$, namely a rotation by $\sigma$ about
the $z$ axis.}
\begin{equation}\label{eq6}
\begin{split}
\sin u &= \sin u' \cos\sigma
+ \cos u' \cos\alpha' \sin\sigma,\\
- \cos u\cos\alpha &= -\sin u' \sin\sigma
+ \cos u' \cos\alpha' \cos\sigma,\\
-\cos u\sin\alpha &= \cos u'\sin\alpha'.
\end{split}
\end{equation}
It is convenient to write these in terms of the auxiliary angles $m$ and
$M$ defined by\footnote
{The auxiliary angles $m$ and $M$ are an angle and a side of the
spherical triangle $EAN$ shown in Fig.\ \ref{figtri}.  Equations
(\ref{eq7}) are the sine rule on angles $E$ and $F$ of triangle $AEF$,
the cosine rule on angle $F$ of triangle $AEF$, and the sine rule on
angles $A$ and $E$ of triangle $ANE$.}
\begin{equation}\label{eq7}
\begin{split}
\sin u' &= \cos m\sin M,\\
\cos u'\cos\alpha' &= \cos m\cos M,\\
\cos u'\sin\alpha' &= \sin m.
\end{split}
\end{equation}
Equations (\ref{eq6}) then become\footnote
{These are analogs of Eqs.~(\ref{eq7}) with meridian $N\!AF$ replaced by
$N\!BG$.}
\begin{equation}\label{eq8}
\begin{split}
\sin u &= \cos m\sin (M+\sigma),\\
\cos u\cos\alpha &= -\cos m\cos (M+\sigma),\\
\cos u\sin\alpha &= -\sin m.
\end{split}
\end{equation}
This gives
\[
\cos^2u = 1 - \cos^2m\sin^2(M+\sigma),
\]
and the equation for $\d s$ becomes
\begin{equation}\label{eq9}
\d s = a \sqrt{1 - e^2}\sqrt{1 + k^2\sin^2(M+\sigma)}\,\d\sigma,
\end{equation}
where
\[
k = \frac{e\cos m}{\sqrt{1-e^2}}.
\]
This differential equation may be integrated in terms of the elliptic
integrals introduced by Legendre.\footnote
{A. M. Legendre,
\href{http://books.google.com/books?id=riIOAAAAQAAJ&printsec=titlepage}
{\it Exercices du calcul int\'egral}, Vol.~1 (Courcier,
1811).}
Because the tools to compute these special functions are not yet
sufficiently versatile,
we instead develop a series solution which converges rapidly because
$e^2$ is so small.  We readily achieve this by decomposing the term
under the square root into two complex factors, namely\footnote
{The notation has been simplified here compared to Bessel's original
formulation in which $k$ and $\epsilon$ are expressed in terms of $E$
through $k = \tan E$ and $\epsilon = \tan^2\frac12E$.
By using $\epsilon$ as the expansion parameter and by dividing out the
factor $1-\epsilon$, Bessel has ensured that the terms that he is
expanding are invariant under the transformation $\epsilon \rightarrow
-\epsilon$, $M+\sigma \rightarrow \pi/2 - (M + \sigma)$.  This symmetry
causes half the terms in the expansions in $\epsilon$ to vanish.}
\begin{multline*}
\d s =
a\frac{\sqrt{1-e^2}}{1-\epsilon}\,\d\sigma \times\\
\sqrt{1 - \epsilon \exp\bigl(2i(M+\sigma)\bigr)}
\sqrt{1 - \epsilon \exp\bigl(-2i(M+\sigma)\bigr)}\,,
\end{multline*}
where
\[
\epsilon = \frac{\sqrt{1+k^2}-1}{\sqrt{1+k^2}+1},\quad
k = \frac{2\sqrt\epsilon}{1-\epsilon}.
\]
Expanding the two factors in the radicals in infinite series
and multiplying the results gives\footnote
{\label{footexp}The use of complex exponentials facilitates the series
expansions by avoiding the need to employ awkward trigonometric
identities.  If we write
$\sqrt{1-x} = 1
- \frac12x
- \frac{1\cdot1}{2\cdot4}x^2
- \frac{1\cdot1\cdot3}{2\cdot4\cdot6}x^3
- \frac{1\cdot1\cdot3\cdot5}{2\cdot4\cdot6\cdot8}x^4
- \ldots = \sum_j a_jx^j$, then the coefficient of
$\cos\bigl(2l(M+\sigma)\bigr)\epsilon^{l+2j}$ is $a_j^2$ for
$l = 0$ and $2a_ja_{j+l}$ for $l > 0$.}
\begin{multline*}
\d s = a
\frac{\sqrt{1-e^2}}{1-\epsilon}
\,
\d\sigma\bigl[A - 2B\cos 2(M+\sigma) \\
- 2C\cos 4(M+\sigma) - 2D\cos 6(M+\sigma) - \ldots\bigr],
\end{multline*}
where $A$, $B$, $C$, \ldots\ are given by
\begin{align*}
A & = 1 + \biggl(\frac12\biggr)^2\epsilon^2
 + \biggl(\frac{1\p1}{2\p4}\biggr)^2\epsilon^4
 + \biggl(\frac{1\p1\p3}{2\p4\p6}\biggr)^2\epsilon^6 + \ldots,\\
B & = \frac12\epsilon
 - \frac{1\p1}{2\p4}\,\frac12\epsilon^3
 - \frac{1\p1\p3}{2\p4\p6}\,\frac{1\p1}{2\p4}\epsilon^5\\
&\qquad\qquad\qquad\qquad
 - \frac{1\p1\p3\p5}{2\p4\p6\p8}\,
   \frac{1\p1\p3}{2\p4\p6}\epsilon^7
 - \ldots,\\
C & = \frac{1\p1}{2\p4}\epsilon^2
 - \frac{1\p1\p3}{2\p4\p6}\,\frac12\epsilon^4
 - \frac{1\p1\p3\p5}{2\p4\p6\p8}\,
   \frac{1\p1}{2\p4}\epsilon^6\\
&\qquad\qquad\qquad\qquad
 - \frac{1\p1\p3\p5\p7}{2\p4\p6\p8\p10}\,
   \frac{1\p1\p3}{2\p4\p6}\epsilon^8 - \ldots,\\
D & = \frac{1\p1\p3}{2\p4\p6}\epsilon^3
 - \frac{1\p1\p3\p5}{2\p4\p6\p8}\,\frac12\epsilon^5
 - \frac{1\p1\p3\p5\p7}{2\p4\p6\p8\p10}\,
   \frac{1\p1}{2\p4}\epsilon^7\\
&\qquad\qquad\qquad\qquad
 - \frac{1\p1\p3\p5\p7\p9}{2\p4\p6\p8\p10\p12}\,
   \frac{1\p1\p3}{2\p4\p6}\epsilon^9 - \ldots,\\
&\mathrm{etc.}
\end{align*}
Integrating the equation for $\d s$ starting at $\sigma = 0$, we obtain
\begin{align}
s =
\frac b{1-\epsilon}
\bigl[A\sigma &- {\textstyle\frac21}B\cos(2M+\sigma)\sin\sigma
\notag\\
&- {\textstyle\frac22}C\cos(4M+2\sigma)\sin2\sigma\notag\\
&- {\textstyle\frac23}D\cos(6M+3\sigma)\sin3\sigma\notag\\
&- \ldots\bigr].\label{eq10}
\end{align}

\section{Solving the distance equation} \label{sec6}

The series (\ref{eq10}) gives the distance $s$ between $A$ and $B$ in
terms of $u'$, $\alpha'$, and $\sigma$; if, however, $s$ and $\alpha'$
have been measured and $u'$ is known from the latitude at $A$, then
$\sigma$ is obtained by solving (\ref{eq10}).  The latitude of $B$ and
the azimuth of the geodesic there are found from (\ref{eq8}).  Equation
(\ref{eq10}) can be solved either by reverting the series or by
successive approximation---the latter way is however the simplest if the
tables I have compiled are used.

I write\footnote
{The units for $\sigma$, $\alpha$, $\beta$, \ldots\ are arc seconds.
Bessel here adopts a conflicting notation for the coefficient $\alpha$
which should not be confused with the azimuth.}
\begin{multline}\label{eq11}
\sigma = \frac\alpha b s + \beta\cos(2M+\sigma)\sin\sigma
+ \gamma\cos(4M+2\sigma)\sin2\sigma\quad\\
+ \delta\cos(6M+3\sigma)\sin3\sigma
 + \ldots,
\end{multline}
where
\begin{equation*}
\begin{split}
\alpha & = \frac{648\,000}\pi\,
\frac{1-\epsilon}A,\\
\beta & = \frac{648\,000}\pi\,\frac{2B}A,\\
\gamma & = \frac{648\,000}\pi\,\frac{C}A,\\
\delta & = \frac{648\,000}\pi\,\frac{2D}{3A},\\
&\mathrm{etc.}
\end{split}
\end{equation*}
The tables give the logarithms\footnote
{In this paper, $\log x$ denotes the common logarithm (base 10) and we
use $\colog x = \log(1/x)$.  The tables in the original paper contained
a number of errors of one unit in the last place.  These errors do not,
for the most part, affect the results obtained from the tables when
rounded to $0.001''$.  In addition, there were systematic errors in the
tabulated values of $\log\beta$ equivalent to a relative error of order
$\epsilon^2$ in $\beta$ which result in discrepancies from 1 to 17 units
in the last place on the final page (the 6-figure portion) of the
tables.  In calculations involving logarithms, a bar over a numeral
indicates that that numeral should be negated, e.g., $\log0.02 \approx
\b2.3 = (-2) + 0.3$.  In the original paper, logarithms are written
modulo 10, e.g., $\log0.02\approx8.3$.  The notation ``$(-)$'' in these
calculations indicates that the quantity whose logarithm is being taken
is negative.}
of $\alpha$, $\beta$, and $\gamma$ as a function of the argument
\[
\log k= \log\frac{e\cos m}{\sqrt{1-e^2}}.
\]
By this choice, the variation of $\log\beta$ and $\log\gamma$ are very
close to two and four times that of the argument, which simplifies
interpolation into the table.\footnote
{The columns headed $\Delta$ give the first differences of the
immediately preceding columns and aid in interpolating the data.  Bessel
would have used a table of ``proportional parts'' to compute the
interpolated values.}

We take $\alpha s/b$ as the first approximation of $\sigma$, substitute
this into the second term to obtain a second approximation, with which
we recalculate the second term and add the third.  The convergence of
the series is sufficiently fast that, even if the argument is $\b1.1$
(which is only possible if the flattening of the ellipsoid, $1-b/a$,
exceeds $\frac1{128}$), the approximation never needs to be carried
further in order to keep the errors in $\sigma$ under $0.001''$.  The
term involving $\delta$ does not exceed $0.0005''$ at this value of
the argument.

\section{Accuracy of the tables} \label{sec7}

The values of $\log\alpha$ in the table are given to 8 decimal
places.
An error of half a unit in the last place results in an error of only
$0.0005''$ or $0.008\,\mathrm{toise}$ over a distance corresponding to
$\sigma = 12^\circ4'$ or 700\,000 toises.\footnote
{The toise was a French unit of length.  It can be converted to meters
by $1\,\mathrm{toise} = 864\,\mathrm{ligne}$, $443.296\,\mathrm{ligne} =
1\,\mathrm m$, or $1\,\mathrm{toise}\approx 1.949\,\mathrm m$.}
Similarly, I retain only sufficient digits in the tabulation of
$\log\beta$ to ensure that the error in this term is less than
$0.0005''$; for this purpose, I use 6 digits at the end of the table and
fewer digits for smaller values of the argument.  The third term never
exceeds $0.17''$, even at the end of the table; therefore I include only
3 decimal places for $\log\gamma$.  Thus the errors are $0.001''$ for
distances up to 700\,000 toises; even if the distance is of the order of
a quarter meridian (i.e., $\sigma = 90^\circ$), the error is less than
$0.01''$.

\section{An example} \label{sec8}

In order to illustrate the use of the tables, I consider the results
from the great survey by von M\"uffling.\footnote
{F. C. F. von M\"uffling,
\href{http://adsabs.harvard.edu/full/1823AN......2...33V}
{Astron.\ Nachr.\ {\bf 2}(27), 33--38 (1824)}.}
Relative to Seeberg (point $A$), the distance and azimuth to Dunkirk
(point $B$) are\footnote
{Seeberg: $50^\circ56'\mathrm N\; 10^\circ44'\mathrm E$;
Dunkirk: $51^\circ2'\mathrm N\; 2^\circ23'\mathrm E$.}
\begin{equation*}
\begin{split}
\log s &= 5.478\,303\,14,\\
\alpha' &= \dms274:21:3.18:.
\end{split}
\end{equation*}
I assume the latitude of the Observatory at Seeberg to be
$\phi'=\dms50:56:6.7:$
and the ellipsoid parameters to be $\log b = 6.513\,354\,64$, $\log e =
\b2.905\,4355$.\footnote
{In present-day units, this is $a \approx 6377\,\mathrm{km}$, flattening
$f \approx 1/308.6$, $s\approx 586\,\mathrm{km}$.  In this example,
Bessel uses the toise as the unit of length and the second as the unit
of arc.}

From $\tan u' = \sqrt{1-e^2}\tan\phi'$, we find
\begin{align*}
\log\tan\phi' &= 0.090\,626\,65\\
\log\sqrt{1-e^2} &=\b1.998\,590\,60\\
\log\tan u' &=\sumbar{0.089\,217\,25}; & u' &= \dms50:50:39.057:.
\end{align*}
Given $u'$ and $\alpha'$, we can compute $M$, $\cos m$ and $\sin m$
from equations (\ref{eq7}):\footnote
{Bessel solves 3 equations (\ref{eq7}) for 2 unknowns $M$ and $m$.  The
redundancy serves as a check for the hand calculation and can also
improve the accuracy of the calculation, for example, in the case where
$\sin m \approx 1$.}
\begin{align*}
\log\sin u' &= \b1.889\,543\,51\\
\log\cos u' &= \b1.800\,326\,27\\
\log\cos\alpha'&= \b2.880\,037\,33\\
\log\sin\alpha'&= \b1.998\,746\,62(-)\\
\log(\cos m\sin M) &= \sumbar{\b1.889\,543\,51}\\
\log(\cos m\cos M) &= \b2.680\,363\,60\\
\log\,\sin m &= \b1.799\,072\,89(-)\\
M &= \sumbar{\dms86:27:53.949:};&\!\!2M&=\dms172:55:47.9:\\
\log\cos m &= \b1.890\,370\,63 &\!\!4M&=\dms345:51:36:.
\end{align*}
The argument in the tables,
$\log\bigl((e/\sqrt{1-e^2})\cos m\bigr)$, is
\begin{align*}
\log\frac{e}{\sqrt{1-e^2}} &= \b2.906\,845\\
\log\cos m &= \b1.890\,371 \\
\mathrm{Argument} &= \sumbar{\b2.797\,216}.
\end{align*}
Looking up $\log\alpha$ in the tables, and calculating $\alpha s/b$
gives\footnote
{It is necessary to use second differences when interpolating in the
table for $\log\alpha$.  The argument, $\b2.797\,216$, lies $q=0.7216$
of the way between $\b2.79$ and $\b2.80$.  Bessel's central
2nd-order interpolation formula for the last 6 digits of $\log\alpha$
gives
$401\,284+q(-1941)+\frac14q(q-1)(1853-1004-1028) = 399\,892$.
For the other table look-ups, linear interpolation using first
differences suffices.}
\begin{align*}
\log\alpha &= 5.313\,998\,92\\
\colog b &= \b7.486\,645\,36\\
\log s &= 5.478\,303\,14\\
\log \frac{\alpha s} b &= \sumbar{4.278\,947\,42};
& \frac\alpha b s &= \dms5:16:48.481:.
\end{align*}
Adopting this as the first approximation to the value of $\sigma$, we
obtain the second by adding the next term in the series (\ref{eq11}),
\begin{align*}
\log\beta &= 2.305\,94\\
\log\cos(2M+\sigma) &=\b1.999\,79(-)\\
\log\sin\sigma &= \b2.963\,91\\
&\hphantom{=\;\;}\sumbar{1.269\,64(-)} = -18.61''.
\end{align*}
We now update the value of this term with the second approximation of
$\sigma =\dms5:16:48.5:-18.6''=\dms5:16:29.9:$ and so obtain as the
third approximation:
\begin{align*}
\log\beta &= 2.305\,94\\
\log\cos(2M+\sigma) &=\b1.999\,79(-)\\
\log\sin\sigma &= \b2.963\,48\\
&\hphantom{=\;\;}\sumbar{1.269\,21(-)} = -18.587'',
\end{align*}
\begin{align*}
\log\gamma &= \b2.394\\
\log\cos(4M+2\sigma) &=\b1.999\\
\log\sin2\sigma &= \b1.263\\
&\hphantom{=\;\;}\sumbar{\b3.656} = +0.005''.
\end{align*}
Gathering the terms in (\ref{eq11}) gives
$\sigma = \dms5:16:48.481:-18.587''+0.005'' =\dms5:16:29.899:$ and so,
finally, we determine $\alpha$, $u$ and $\phi$ from equations
(\ref{eq8}),
\begin{align*}
M + \sigma &= \dms91:44:23.848:\\
\log\sin(M+\sigma) &= \b1.999\,799\,71\\
\log\bigl(-\cos(M+\sigma)\bigr) &= \b2.482\,349\,32\\
\log\cos m &= \b1.890\,370\,63\\
\log(-\sin m) &= \b1.799\,072\,89\\
\log\,\sin u &= \sumbar{\b1.890\,170\,34}\\
\log(\cos u\cos\alpha) &= \b2.372\,719\,95\\
\log(\cos u\sin\alpha) &= \b1.799\,072\,89\\
\log\cot\alpha &= \sumbar{\b2.573\,647\,06};&
\!\!\!\!\alpha &= \dms87:51:15.523:\\
\log\cos u &= \b1.799\,377\,50\\
\log\tan u &= 0.090\,792\,84\\
\colog\sqrt{1-e^2} &= 0.001\,409\,40\\
\log\tan\phi &= \sumbar{0.092\,202\,24};&
\!\!\!\!\phi&=\dms51:2:12.719:.
\end{align*}
In this example, I carried out the trigonometric calculations to 8
decimals; however the tables of $\log\alpha$, $\log\beta$, and
$\log \gamma$ in fact allow $\alpha$ and $\phi$ to be determined
slightly more accurately than this.  If only standard 7-figure logarithm
tables are available, the last digits in the tabulated values of
$\log\alpha$, $\log\beta$, and $\log\gamma$ may be neglected.

\section{The longitude integral} \label{sec9}

We turn now to the determination of the longitude difference $w$ by
integrating (\ref{eq5}),
\[
\d w = \sqrt{1-e^2\cos^2u}\,\d\omega.
\]
This integral contains two separate constants $m$ and $e$, which cannot
be combined.  Thus it is not possible to construct tables to allow a
rigorous solution of this problem which are valid for arbitrary
$e$.\footnote
{As a practical matter, it would have been impossible for Bessel to
provide a complete tabulation of a function of two parameters.  He could
have tabulated the function for a fixed value of $e$, which would
greatly reduced the utility of his method, especially given the
uncertainties in the measurements of $e$.  Instead, Bessel manipulates
the expression for $\d w$ to move the dependence on the second parameter
into a small term that may be neglected.}
However, we can achieve our goal by sacrificing strict rigor and by
making an approximation which results in errors which are
inconsequential in our application.

We start by writing
\[
\d w = \d\omega - \bigl(1 - \sqrt{1-e^2\cos^2u}\bigr)\d\omega,
\]
and substitute in the second term
\[
\d\omega =\frac{\sin\alpha'\cos u'}{\cos^2u}\,\d\sigma.
\]
On integrating, we obtain
\[
w = \omega - \sin\alpha'\cos u'
\int\frac{1-\sqrt{1-e^2\cos^2u}}{\cos^2u}\,\d\sigma.
\]
Let us write
\[
\frac{1-\sqrt{1-e^2\cos^2u}}{\cos^2u} =
\frac{e^2}2 (1+e^2 p \cos^2 u)^q(1+y);
\]
in other words, we set
\begin{multline*}
1+y = \frac{2(1-\sqrt{1-e^2\cos^2u})}{e^2\cos^2u(1+e^2p\cos^2u)^q}\\
=
\frac{1+\frac14e^2\cos^2u+\frac18e^4\cos^4u+\frac5{64}e^6\cos^6u+\ldots}
{\Biggl(
\begin{aligned}
\textstyle1+qpe^2\cos^2u
+\frac{q(q-1)}{1\cdot2}p^2e^4\cos^4u\qquad\qquad\\
\textstyle\qquad\qquad
+\frac{q(q-1)(q-2)}{1\cdot2\cdot3}p^3e^6\cos^6u+\ldots
\end{aligned}\Biggr)
}.
\end{multline*}
The first three terms in the denominator and in the numerator are equal,
provided that
\[
p = -{\textstyle\frac34}, \qquad q = -{\textstyle\frac13},
\]
which gives
\begin{align*}
1+y &=
\frac{1+\frac14e^2\cos^2u+\frac18e^4\cos^4u+\frac5{64}e^6\cos^6u+\ldots}
{1+\frac14e^2\cos^2u+\frac18e^4\cos^4u+\frac7{96}e^6\cos^6u+\ldots}\\
&=1 + {\textstyle\frac1{192}}e^6\cos^6u + \ldots
\end{align*}
From this, we see that neglecting $y$ results in an error of order $e^8$
or an error in $w$ of $\frac1{384}e^8\sigma$.  This would not be
discernible even in the calculation of long geodesics to 10 decimal
places.\footnote
{For a flattening of $\frac1{128}$, the error in the longitude
difference over a distance equivalent to a quarter meridian, i.e.,
$10\,000\,\mathrm{km}$, is less than $0.000\,05''$.}

Thus, for the present purposes, we may take $y \approx 0$ enabling us
to tabulate the integral in a way that is valid for all $e$.

\section{Series expansion for longitude} \label{sec10}

Introducing this approximation, we have
\begin{align*}
w &\approx \omega - \frac{e^2}2\sin m\int\frac{\d\sigma}
{\sqrt[3]{1-\frac34e^2\cos^2u}}\\
  &= \omega - \frac{e^2}2\sin m\!\int\!\!\frac{\d\sigma}
{\sqrt[3]{1-\frac34e^2
+\frac34e^2\cos^2m\sin^2(M+\sigma)}}.
\end{align*}
If we set
\[
k' = \frac
{\sqrt{{\scriptstyle\frac34}}e\cos m}
{\sqrt{1-{\scriptstyle\frac34}e^2}},
\]
we can express the integral in the second term as
\[
\int\frac{\d\sigma}
{\sqrt[3]{1-\frac34e^2}
\sqrt[3]{1+k'^2\sin^2(M+\sigma)}}.
\]
Following the same procedure used in expanding the integral for $\d s$ in
Sec.~\ref{sec5}, we introduce $\epsilon'$ defined by\footnote
{Bessel gives the relationship between $k'$ and $\epsilon'$ in terms of
$E'$, where $k' = \tan E'$ and $\epsilon' = \tan^2\frac12E'$.}
\[
\epsilon' = \frac{\sqrt{1+k'^2}-1}{\sqrt{1+k'^2}+1},
\quad k' = \frac{2\sqrt{\epsilon'}}{1-\epsilon'},
\]
and separate the integrand into two complex factors,
\[
\int\frac{\sqrt[3]{(1- \epsilon')^2/\bigl(1-\frac34e^2\bigr)}\,\d\sigma}
{\sqrt[3]{1 - \epsilon' \exp\bigl(2i(M+\sigma)\bigr)}
\sqrt[3]{1 - \epsilon' \exp\bigl(-2i(M+\sigma)\bigr)}}
.
\]
If we expand these in infinite series, the product becomes\footnote
{There are a series of errors in the original paper leading up to
(\ref{eq12}).  Here we assume that the original Eq.~(\ref{eq12}) defines
$\alpha'$, $\beta'$, $\gamma'$, \ldots, which makes this equation
analogous to (\ref{eq11}), and correct the preceding equations to be
consistent.}
\begin{multline*}
\!\!\!\!\frac2{\sqrt[3]{1-{\scriptstyle\frac34}e^2}}
\int\Bigl(
\alpha'
+\beta'\cos2(M+\sigma)
+2\gamma'\cos4(M+\sigma)\\
{}+3\delta'\cos6(M+\sigma)+\ldots
\Bigr)\d\sigma,
\end{multline*}
where\footnote
{See footnote \ref{footexp} and set
$(1-x)^{-1/3} = 1
+ \frac13x
+ \frac{1\cdot4}{3\cdot6}x^2
+ \frac{1\cdot4\cdot7}{3\cdot6\cdot9}x^3
+ \frac{1\cdot4\cdot7\cdot10}{3\cdot6\cdot9\cdot12}x^4
+ \ldots$}
\begin{align*}
\alpha'&={\textstyle\frac12}
\sqrt[3]{(1- \epsilon')^2}\biggl[1
+ \biggl(\frac13\biggr)^2\epsilon'^2
+ \biggl(\frac{1\p4}{3\p6}\biggr)^2\epsilon'^4
+\ldots\biggr],\\
\beta'&={\textstyle\frac11}
\sqrt[3]{(1- \epsilon')^2}\biggl[\frac13\epsilon'
+ \frac{1\p4}{3\p6}\,\frac13\epsilon'^3
+ \frac{1\p4\p7}{3\p6\p9}\,\frac{1\p4}{3\p6}\epsilon'^5
+\ldots\biggr],\\
\gamma'&={\textstyle\frac12}
\sqrt[3]{(1- \epsilon')^2}\biggl[\frac{1\p4}{3\p6}\epsilon'^2
+ \frac{1\p4\p7}{3\p6\p9}\,\frac13\epsilon'^4\\
&\qquad\qquad\qquad\qquad\qquad\quad
+ \frac{1\p4\p7\p10}{3\p6\p9\p12}\,\frac{1\p4}{3\p6}\epsilon'^6
+\ldots\biggr],\\
\delta'&={\textstyle\frac13}
\sqrt[3]{(1- \epsilon')^2}\biggl[\frac{1\p4\p7}{3\p6\p9}\epsilon'^3
+ \frac{1\p4\p7\p10}{3\p6\p9\p12}\,\frac13\epsilon'^5\\
&\qquad\qquad\qquad\qquad\qquad\quad
+ \frac{1\p4\p7\p10\p13}{3\p6\p9\p12\p15}\,\frac{1\p4}{3\p6}\epsilon'^7
+\ldots\biggr],\\
&\mathrm{etc.}
\end{align*}
Integrating from $\sigma = 0$ then gives
\begin{align}\label{eq12}
w &\approx \omega - \frac{e^2\sin m}{\sqrt[3]{1-{\scriptstyle\frac34}e^2}}
\Bigl(\alpha'\sigma + \beta' \cos(2M+\sigma)\sin\sigma\notag\\
&\qquad\qquad\qquad+\gamma' \cos(4M+2\sigma)\sin2\sigma\notag\\
&\qquad\qquad\qquad+\delta' \cos(6M+3\sigma)\sin3\sigma + \ldots\Bigr)
.
\end{align}

\section{Computing the longitude difference} \label{sec11}

The first two coefficients of this series are given in the 4th and 5th
columns of the tables\footnote
{The value of $\beta'$ in the tables includes the factor of
$648\,000/\pi$ necessary to convert from radians to arc seconds.}  as
functions of the argument
\[
\log k'=
\log\biggl(\frac{\sqrt{{\scriptstyle\frac34}}e}
{\sqrt{1-{\scriptstyle\frac34}e^2}}\cos m\biggr).
\]
The convergence is commensurate with the 3 first columns of the
tables.  We calculate $\omega$ using one of the formulas for spherical
triangles (Sec.~\ref{sec3}), either\footnote
{The first two relations are the sine rule for angle $N$ of
triangle $ABN$ of Fig.~\ref{figtri}.  The last relation is obtained,
for example, by substituting for $\sin\alpha'$ from (\ref{eq7}).}
\[
\sin \omega = \frac{\sin\sigma\sin\alpha'}{\cos u}
= \frac{-\sin\sigma\sin\alpha}{\cos u'}
= \frac{\sin\sigma\sin m}{\cos u \cos u'},
\]
or\footnote
{These are Napier's analogies for angle $N$ of triangle $ABN$.}
\begin{align*}
\tan {\textstyle\frac12}\omega &=
\frac{\sin\frac12(u'-u)}{\cos\frac12(u'+u)}
\cot{\textstyle\frac12}(\alpha'+\alpha)\\
&=
\frac{\cos\frac12(u'-u)}{\sin\frac12(u'+u)}
\cot{\textstyle\frac12}(\alpha'-\alpha).
\end{align*}
and evaluate $w$ by means of the tables.

I will continue with the example in Sec.~\ref{sec8} and calculate the
longitude difference between Dunkirk and Seeberg using this
prescription.  Solving the spherical triangle for $\omega$ gives
\begin{align*}
\log\sin\sigma &= \b2.963\,483\,83\\
\log(-\sin\alpha) &= \b1.999\,695\,39(-)\\
\colog\cos u' &= 0.199\,673\,73\\
\log\sin\omega &= \sumbar{\b1.162\,852\,95}(-);
&\omega &= -\dms8:21:57.741:.
\end{align*}
The argument for the last two columns of the tables is
$\log\bigl((\sqrt{\scriptstyle\frac34}e/
\sqrt{1-{\scriptstyle\frac34}e^2})\cos m\bigr)$, giving
\begin{align*}
\log\frac{\sqrt{{\scriptstyle\frac34}}e}
{\sqrt{1-{\scriptstyle\frac34}e^2}} &= \b2.844\,022\\
\log\cos m &= \b1.890\,371\\
\mathrm{Argument} &= \b2.734\,393.
\end{align*}
Computing the terms in the series (\ref{eq12}) gives
\begin{align*}
\log\alpha' &= \b1.698\,758\\
\log(-\sin m) &= \b1.799\,073\\
\log\frac{e^2}{\sqrt[3]{1-{\scriptstyle\frac34}e^2}} &= \b3.811\,575\\
\log\sigma &=4.278\,523\\
&\hphantom{=\;\;}\sumbar{1.587\,929} = +38.719'',
\end{align*}
and
\begin{align*}
\log\beta' &= 1.703\\
\log(-\sin m) &= \b1.799\\
\log\frac{e^2}{\sqrt[3]{1-{\scriptstyle\frac34}e^2}} &= \b3.812\\
\log\bigl(\cos(2M+\sigma)\sin\sigma\bigr) &=\b2.963(-)\\
&\hphantom{=\;\;}\sumbar{\b2.277}(-) = -0.019''.
\end{align*}
The sum of both terms is $+38.700''$, and adding this to $\omega$, we
find the longitude difference,
\[
w = -\dms8:21:19.041:.
\]

\section{Conclusion} \label{sec12}

This illustration of the use of these tables shows that the accuracy of
the calculation is limited not by the neglect of terms of high order in
the eccentricity, but by the number of decimal places included (i.e.,
the truncation error is smaller than the round-off error).  The
steps in the calculation are, for the most part, the same as for a
spherical earth; in order to account for the earth's ellipticity one
needs, in addition, only to solve equation (\ref{eq11}) and to evaluate
the series (\ref{eq12}).  Since this approach is sufficiently convenient
even for routine use, it is unnecessary to use an approximate method
which is valid only for small distances.

\begin{center}
(The tables are shown on the following pages.)
\end{center}


\def\head{\iflanguage{german}
{{\sc Tafeln} zur Berechnung der geod"atischen Vermessungen}
{{\sc Tables} for computing geodesics}}
\setlength{\extrarowheight}{0.8pt}
\def\minus{-}\def\v{\\[0.8ex]}
\def\shim{\vphantom{\bigr|^{2^2}}}
\catcode`\+=\active\def+{}\catcode`\-=\active\let-=\b
\catcode`\,=\active\def,{\,}\catcode`\.=\active\let.=@
\catcode`\?=\active\def?{\hphantom0}
\begin{table*}
\begin{center}
\head, 1.
\small
\begin{tabular}{ | >{\quad$}c<{$\quad} |
>{\quad$}c<{$} >{$\hspace{1em}}r<{$\quad} |
>{\quad$}c<{$} >{$\hspace{1em}}r<{$\quad} |
>{\quad$}c<{$} >{$\hspace{1em}}r<{$\quad} |
>{\quad$}c<{$} >{$\hspace{1em}}r<{$\quad} |
>{\quad$}c<{$} >{$\hspace{1em}}r<{$\quad} |}
\hline\shim
\mathrm{Arg}
&\log\alpha&\hspace{\minus1em}\minus\Delta
&\log\beta&\hspace{\minus1em}\hphantom{\minus}\Delta
&\log\gamma\quad&\hspace{\minus1em}\hphantom{\minus}\Delta
&\log\alpha'&\hspace{\minus1em}\minus\Delta
&\log\beta'&\hspace{\minus1em}\hphantom{\minus}\Delta
\\[\extrarowheight]\hline\shim
-4.4? &5.314,425,13&   1&-3.5124   &2000&      &  &-1.698,970& 0&-3.035&200\v
-4.5? &5.314,425,12&   0&-3.7124   &2000&      &  &-1.698,970& 0&-3.235&200\\
-4.6? &5.314,425,12&   1&-3.9124   &2000&      &  &-1.698,970& 0&-3.435&200\\
-4.7? &5.314,425,11&   2&-2.1124   &2000&      &  &-1.698,970& 0&-3.635&200\\
-4.8? &5.314,425,09&   3&-2.3124   &2000&      &  &-1.698,970& 0&-3.835&200\\
-4.9? &5.314,425,06&   4&-2.5124   &2000&      &  &-1.698,970& 0&-2.035&200\v
-3.0? &5.314,425,02&   6&-2.7124   &2000&      &  &-1.698,970& 0&-2.235&200\\
-3.1? &5.314,424,96&  10&-2.9124   &2000&      &  &-1.698,970& 0&-2.435&200\\
-3.2? &5.314,424,86&  16&-1.1124   &2000&      &  &-1.698,970& 0&-2.635&200\\
-3.3? &5.314,424,70&  25&-1.3124   &2000&      &  &-1.698,970& 0&-2.835&200\\
-3.4? &5.314,424,45&  40&-1.5124   &2000&      &  &-1.698,970& 1&-1.035&200\v
-3.50 &5.314,424,05&   5&-1.7124   & 200&      &  &-1.698,969& 0&-1.235& 20\\
-3.51 &5.314,424,00&   6&-1.7324   & 200&      &  &-1.698,969& 0&-1.255& 20\\
-3.52 &5.314,423,94&   5&-1.7524   & 200&      &  &-1.698,969& 0&-1.275& 20\\
-3.53 &5.314,423,89&   6&-1.7724   & 200&      &  &-1.698,969& 0&-1.295& 20\\
-3.54 &5.314,423,83&   6&-1.7924   & 200&      &  &-1.698,969& 0&-1.315& 20\v
-3.55 &5.314,423,77&   7&-1.8124   & 200&      &  &-1.698,969& 0&-1.335& 20\\
-3.56 &5.314,423,70&   7&-1.8324   & 200&      &  &-1.698,969& 0&-1.355& 20\\
-3.57 &5.314,423,63&   7&-1.8524   & 200&      &  &-1.698,969& 0&-1.375& 20\\
-3.58 &5.314,423,56&   7&-1.8724   & 200&      &  &-1.698,969& 0&-1.395& 20\\
-3.59 &5.314,423,49&   8&-1.8924   & 200&      &  &-1.698,969& 0&-1.415& 20\v
-3.60 &5.314,423,41&   8&-1.9124   & 200&      &  &-1.698,969& 0&-1.435& 20\\
-3.61 &5.314,423,33&   8&-1.9324   & 200&      &  &-1.698,969& 0&-1.455& 20\\
-3.62 &5.314,423,25&   9&-1.9524   & 200&      &  &-1.698,969& 0&-1.475& 20\\
-3.63 &5.314,423,16&  10&-1.9724   & 200&      &  &-1.698,969& 0&-1.495& 20\\
-3.64 &5.314,423,06&   9&-1.9924   & 200&      &  &-1.698,969& 0&-1.515& 20\v
-3.65 &5.314,422,97&  11&+0.0124   & 200&      &  &-1.698,969& 1&-1.535& 20\\
-3.66 &5.314,422,86&  10&+0.0324   & 200&      &  &-1.698,968& 0&-1.555& 20\\
-3.67 &5.314,422,76&  11&+0.0524   & 200&      &  &-1.698,968& 0&-1.575& 20\\
-3.68 &5.314,422,65&  12&+0.0724   & 200&      &  &-1.698,968& 0&-1.595& 20\\
-3.69 &5.314,422,53&  12&+0.0924   & 200&      &  &-1.698,968& 0&-1.615& 20\v
-3.70 &5.314,422,41&  13&+0.1124   & 200&      &  &-1.698,968& 0&-1.635& 20\\
-3.71 &5.314,422,28&  14&+0.1324   & 200&      &  &-1.698,968& 0&-1.655& 20\\
-3.72 &5.314,422,14&  14&+0.1524   & 200&      &  &-1.698,968& 0&-1.675& 20\\
-3.73 &5.314,422,00&  15&+0.1724   & 200&      &  &-1.698,968& 0&-1.695& 20\\
-3.74 &5.314,421,85&  15&+0.1924   & 200&      &  &-1.698,968& 0&-1.715& 20\v
-3.75 &5.314,421,70&  16&+0.2124   & 200&      &  &-1.698,968& 0&-1.735& 20\\
-3.76 &5.314,421,54&  17&+0.2324   & 200&      &  &-1.698,968& 1&-1.755& 20\\
-3.77 &5.314,421,37&  18&+0.2524   & 200&      &  &-1.698,967& 0&-1.775& 20\\
-3.78 &5.314,421,19&  18&+0.2724   & 200&      &  &-1.698,967& 0&-1.795& 20\\
-3.79 &5.314,421,01&  20&+0.2924   & 200&      &  &-1.698,967& 0&-1.815& 20\v
-3.80 &5.314,420,81&  20&+0.3124   & 200&      &  &-1.698,967& 0&-1.835& 20\\
-3.81 &5.314,420,61&  22&+0.3324   & 200&      &  &-1.698,967& 0&-1.855& 20\\
-3.82 &5.314,420,39&  22&+0.3524   & 200&      &  &-1.698,967& 0&-1.875& 20\\
-3.83 &5.314,420,17&  23&+0.3724   & 200&      &  &-1.698,967& 0&-1.895& 20\\
-3.84 &5.314,419,94&  25&+0.3924   & 200&      &  &-1.698,967& 1&-1.915& 20\v
-3.85 &5.314,419,69&  25&+0.4124   & 200&      &  &-1.698,966& 0&-1.935& 20\\
-3.86 &5.314,419,44&  27&+0.4324   & 200&      &  &-1.698,966& 0&-1.955& 20\\
-3.87 &5.314,419,17&  28&+0.4524   & 200&      &  &-1.698,966& 0&-1.975& 20\\
-3.88 &5.314,418,89&  30&+0.4724   & 200&      &  &-1.698,966& 0&-1.995& 20\\
-3.89 &5.314,418,59&  31&+0.4924   & 200&      &  &-1.698,966& 1&+0.015& 20\v
-3.90 &5.314,418,28&    &+0.5124   &    &      &  &-1.698,965&  &+0.035&   \\
[\extrarowheight]\hline
\end{tabular}
\end{center}
\end{table*}
\begin{table*}
\begin{center}
\head, 2.
\small
\begin{tabular}{ | >{\quad$}c<{$\quad} |
>{\quad$}c<{$} >{$\hspace{1em}}r<{$\quad} |
>{\quad$}c<{$} >{$\hspace{1em}}r<{$\quad} |
>{\quad$}c<{$} >{$\hspace{1em}}r<{$\quad} |
>{\quad$}c<{$} >{$\hspace{1em}}r<{$\quad} |
>{\quad$}c<{$} >{$\hspace{1em}}r<{$\quad} |}
\hline\shim
\mathrm{Arg}
&\log\alpha&\hspace{\minus1em}\minus\Delta
&\log\beta&\hspace{\minus1em}\hphantom{\minus}\Delta
&\log\gamma\quad&\hspace{\minus1em}\hphantom{\minus}\Delta
&\log\alpha'&\hspace{\minus1em}\minus\Delta
&\log\beta'&\hspace{\minus1em}\hphantom{\minus}\Delta
\\[\extrarowheight]\hline\shim
-3.90 &5.314,418,28&  32&+0.512,35 &2000&      &  &-1.698,965& 0&+0.035& 20\\
-3.91 &5.314,417,96&  34&+0.532,35 &2000&      &  &-1.698,965& 0&+0.055& 20\\
-3.92 &5.314,417,62&  35&+0.552,35 &2000&      &  &-1.698,965& 0&+0.075& 20\\
-3.93 &5.314,417,27&  37&+0.572,35 &2000&      &  &-1.698,965& 0&+0.095& 20\\
-3.94 &5.314,416,90&  39&+0.592,35 &2000&      &  &-1.698,965& 1&+0.115& 20\v
-3.95 &5.314,416,51&  41&+0.612,35 &2000&      &  &-1.698,964& 0&+0.135& 20\\
-3.96 &5.314,416,10&  42&+0.632,35 &2000&      &  &-1.698,964& 0&+0.155& 20\\
-3.97 &5.314,415,68&  45&+0.652,35 &2000&      &  &-1.698,964& 1&+0.175& 20\\
-3.98 &5.314,415,23&  47&+0.672,35 &1999&      &  &-1.698,963& 0&+0.195& 20\\
-3.99 &5.314,414,76&  48&+0.692,34 &2000&      &  &-1.698,963& 0&+0.215& 20\v
-2.00 &5.314,414,28&  52&+0.712,34 &2000&      &  &-1.698,963& 1&+0.235& 20\\
-2.01 &5.314,413,76&  53&+0.732,34 &2000&      &  &-1.698,962& 0&+0.255& 20\\
-2.02 &5.314,413,23&  56&+0.752,34 &2000&      &  &-1.698,962& 0&+0.275& 20\\
-2.03 &5.314,412,67&  59&+0.772,34 &2000&      &  &-1.698,962& 1&+0.295& 20\\
-2.04 &5.314,412,08&  61&+0.792,34 &2000&      &  &-1.698,961& 0&+0.315& 20\v
-2.05 &5.314,411,47&  65&+0.812,34 &2000&      &  &-1.698,961& 1&+0.335& 20\\
-2.06 &5.314,410,82&  67&+0.832,34 &2000&      &  &-1.698,960& 0&+0.355& 20\\
-2.07 &5.314,410,15&  71&+0.852,34 &1999&      &  &-1.698,960& 0&+0.375& 20\\
-2.08 &5.314,409,44&  74&+0.872,33 &2000&      &  &-1.698,960& 1&+0.395& 20\\
-2.09 &5.314,408,70&  77&+0.892,33 &2000&      &  &-1.698,959& 0&+0.415& 20\v
-2.10 &5.314,407,93&  81&+0.912,33 &2000&      &  &-1.698,959& 1&+0.435& 20\\
-2.11 &5.314,407,12&  85&+0.932,33 &2000&      &  &-1.698,958& 1&+0.455& 20\\
-2.12 &5.314,406,27&  89&+0.952,33 &2000&      &  &-1.698,957& 0&+0.475& 20\\
-2.13 &5.314,405,38&  93&+0.972,33 &1999&      &  &-1.698,957& 1&+0.495& 20\\
-2.14 &5.314,404,45&  98&+0.992,32 &2000&      &  &-1.698,956& 0&+0.515& 20\v
-2.15 &5.314,403,47& 102&+1.012,32 &2000&      &  &-1.698,956& 1&+0.535& 20\\
-2.16 &5.314,402,45& 107&+1.032,32 &2000&      &  &-1.698,955& 1&+0.555& 20\\
-2.17 &5.314,401,38& 112&+1.052,32 &2000&      &  &-1.698,954& 1&+0.575& 20\\
-2.18 &5.314,400,26& 117&+1.072,32 &1999&      &  &-1.698,953& 0&+0.595& 20\\
-2.19 &5.314,399,09& 123&+1.092,31 &2000&      &  &-1.698,953& 1&+0.615& 20\v
-2.20 &5.314,397,86& 128&+1.112,31 &2000&      &  &-1.698,952& 1&+0.635& 20\\
-2.21 &5.314,396,58& 135&+1.132,31 &2000&      &  &-1.698,951& 1&+0.655& 20\\
-2.22 &5.314,395,23& 141&+1.152,31 &1999&      &  &-1.698,950& 1&+0.675& 20\\
-2.23 &5.314,393,82& 147&+1.172,30 &2000&      &  &-1.698,949& 1&+0.695& 20\\
-2.24 &5.314,392,35& 155&+1.192,30 &2000&      &  &-1.698,948& 1&+0.715& 20\v
-2.25 &5.314,390,80& 162&+1.212,30 &1999&-4.207&40&-1.698,947& 1&+0.735& 20\\
-2.26 &5.314,389,18& 169&+1.232,29 &2000&-4.247&40&-1.698,946& 1&+0.755& 20\\
-2.27 &5.314,387,49& 177&+1.252,29 &2000&-4.287&40&-1.698,945& 1&+0.775& 20\\
-2.28 &5.314,385,72& 186&+1.272,29 &1999&-4.327&40&-1.698,944& 2&+0.795& 20\\
-2.29 &5.314,383,86& 195&+1.292,28 &2000&-4.367&40&-1.698,942& 1&+0.815& 20\v
-2.30 &5.314,381,91& 203&+1.312,28 &1999&-4.407&40&-1.698,941& 1&+0.835& 20\\
-2.31 &5.314,379,88& 213&+1.332,27 &2000&-4.447&40&-1.698,940& 2&+0.855& 20\\
-2.32 &5.314,377,75& 224&+1.352,27 &2000&-4.487&40&-1.698,938& 1&+0.875& 20\\
-2.33 &5.314,375,51& 234&+1.372,27 &1999&-4.527&40&-1.698,937& 2&+0.895& 20\\
-2.34 &5.314,373,17& 244&+1.392,26 &2000&-4.567&40&-1.698,935& 1&+0.915& 20\v
-2.35 &5.314,370,73& 257&+1.412,26 &1999&-4.607&40&-1.698,934& 2&+0.935& 20\\
-2.36 &5.314,368,16& 268&+1.432,25 &2000&-4.647&40&-1.698,932& 2&+0.955& 20\\
-2.37 &5.314,365,48& 281&+1.452,25 &1999&-4.687&40&-1.698,930& 2&+0.975& 20\\
-2.38 &5.314,362,67& 295&+1.472,24 &1999&-4.727&40&-1.698,928& 2&+0.995& 20\\
-2.39 &5.314,359,72& 308&+1.492,23 &2000&-4.767&40&-1.698,926& 2&+1.015& 20\v
-2.40 &5.314,356,64&    &+1.512,23 &    &-4.807&  &-1.698,924&  &+1.035&   \\
[\extrarowheight]\hline
\end{tabular}
\end{center}
\end{table*}
\begin{table*}
\begin{center}
\head, 3.
\small
\begin{tabular}{ | >{\quad$}c<{$\quad} |
>{\quad$}c<{$} >{$\hspace{1em}}r<{$\quad} |
>{\quad$}c<{$} >{$\hspace{1em}}r<{$\quad} |
>{\quad$}c<{$} >{$\hspace{1em}}r<{$\quad} |
>{\quad$}c<{$} >{$\hspace{1em}}r<{$\quad} |
>{\quad$}c<{$} >{$\hspace{1em}}r<{$\quad} |}
\hline\shim
\mathrm{Arg}
&\log\alpha&\hspace{\minus1em}\minus\Delta
&\log\beta&\hspace{\minus1em}\hphantom{\minus}\Delta
&\log\gamma\quad&\hspace{\minus1em}\hphantom{\minus}\Delta
&\log\alpha'&\hspace{\minus1em}\minus\Delta
&\log\beta'&\hspace{\minus1em}\hphantom{\minus}\Delta
\\[\extrarowheight]\hline\shim
-2.40?&5.314,356,64& 323&+1.512,23 &1999&-4.807&40&-1.698,924& 2&+1.035& 20\\
-2.41?&5.314,353,41& 338&+1.532,22 &1999&-4.847&40&-1.698,922& 2&+1.055& 20\\
-2.42?&5.314,350,03& 353&+1.552,21 &2000&-4.887&40&-1.698,920& 2&+1.075& 20\\
-2.43?&5.314,346,50& 371&+1.572,21 &1999&-4.927&40&-1.698,918& 3&+1.095& 20\\
-2.44?&5.314,342,79& 388&+1.592,20 &1999&-4.967&40&-1.698,915& 2&+1.115& 20\v
-2.45?&5.314,338,91& 406&+1.612,19 &1999&-3.007&40&-1.698,913& 3&+1.135& 20\\
-2.46?&5.314,334,85& 425&+1.632,18 &2000&-3.047&40&-1.698,910& 3&+1.155& 20\\
-2.47?&5.314,330,60& 446&+1.652,18 &1999&-3.087&40&-1.698,907& 3&+1.175& 20\\
-2.48?&5.314,326,14& 466&+1.672,17 &1999&-3.127&40&-1.698,904& 3&+1.195& 20\\
-2.49?&5.314,321,48& 489&+1.692,16 &1999&-3.167&40&-1.698,901& 3&+1.215& 20\v
-2.50?&5.314,316,59& 511&+1.712,15 &1999&-3.207&40&-1.698,898& 4&+1.235& 20\\
-2.51?&5.314,311,48& 535&+1.732,14 &1999&-3.247&40&-1.698,894& 3&+1.255& 20\\
-2.52?&5.314,306,13& 561&+1.752,13 &1999&-3.287&40&-1.698,891& 4&+1.275& 20\\
-2.53?&5.314,300,52& 587&+1.772,12 &1998&-3.327&40&-1.698,887& 4&+1.295& 20\\
-2.54?&5.314,294,65& 615&+1.792,10 &1999&-3.367&40&-1.698,883& 4&+1.315& 20\v
-2.55?&5.314,288,50& 644&+1.812,09 &1999&-3.407&40&-1.698,879& 4&+1.335& 20\\
-2.56?&5.314,282,06& 674&+1.832,08 &1999&-3.447&40&-1.698,875& 5&+1.355& 20\\
-2.57?&5.314,275,32& 705&+1.852,07 &1998&-3.487&40&-1.698,870& 5&+1.375& 20\\
-2.58?&5.314,268,27& 739&+1.872,05 &1999&-3.527&40&-1.698,865& 4&+1.395& 20\\
-2.59?&5.314,260,88& 774&+1.892,04 &1998&-3.567&40&-1.698,861& 6&+1.415& 20\v
-2.60?&5.314,253,14& 810&+1.912,02 &1998&-3.607&39&-1.698,855& 5&+1.435& 20\\
-2.61?&5.314,245,04& 848&+1.932,00 &1999&-3.646&40&-1.698,850& 6&+1.455& 20\\
-2.62?&5.314,236,56& 889&+1.951,99 &1998&-3.686&40&-1.698,844& 6&+1.475& 20\\
-2.63?&5.314,227,67& 930&+1.971,97 &1998&-3.726&40&-1.698,838& 6&+1.495& 20\\
-2.64?&5.314,218,37& 973&+1.991,95 &1998&-3.766&40&-1.698,832& 6&+1.515& 20\v
-2.65?&5.314,208,64&1020&+2.011,93 &1998&-3.806&40&-1.698,826& 7&+1.535& 20\\
-2.66?&5.314,198,44&1068&+2.031,91 &1998&-3.846&40&-1.698,819& 7&+1.555& 20\\
-2.67?&5.314,187,76&1118&+2.051,89 &1998&-3.886&40&-1.698,812& 8&+1.575& 20\\
-2.68?&5.314,176,58&1170&+2.071,87 &1997&-3.926&40&-1.698,804& 7&+1.595& 20\\
-2.69?&5.314,164,88&1226&+2.091,84 &1998&-3.966&40&-1.698,797& 9&+1.615& 20\v
-2.70?&5.314,152,62&1283&+2.111,82 &1997&-2.006&40&-1.698,788& 8&+1.635& 19\\
-2.71?&5.314,139,79&1344&+2.131,79 &1998&-2.046&40&-1.698,780& 9&+1.654& 20\\
-2.72?&5.314,126,35&1406&+2.151,77 &1997&-2.086&40&-1.698,771& 9&+1.674& 20\\
-2.73?&5.314,112,29&1473&+2.171,74 &1997&-2.126&40&-1.698,762&10&+1.694& 20\\
-2.74?&5.314,097,56&1543&+2.191,71 &1997&-2.166&40&-1.698,752&11&+1.714& 20\v
-2.75?&5.314,082,13&1615&+2.211,68 &1997&-2.206&40&-1.698,741&10&+1.734& 20\\
-2.76?&5.314,065,98&1690&+2.231,65 &1996&-2.246&40&-1.698,731&12&+1.754& 20\\
-2.77?&5.314,049,08&1771&+2.251,61 &1997&-2.286&40&-1.698,719&11&+1.774& 20\\
-2.78?&5.314,031,37&1853&+2.271,58 &1996&-2.326&40&-1.698,708&13&+1.794& 20\\
-2.79?&5.314,012,84&1941&+2.291,54 &1996&-2.366&39&-1.698,695&13&+1.814& 20\v
-2.800&5.313,993,43&1004&+2.311,50 & 998&-2.405&20&-1.698,682& 6&+1.834& 10\\
-2.805&5.313,983,39&1028&+2.321,48 & 998&-2.425&20&-1.698,676& 7&+1.844& 10\\
-2.810&5.313,973,11&1051&+2.331,46 & 998&-2.445&20&-1.698,669& 7&+1.854& 10\\
-2.815&5.313,962,60&1076&+2.341,44 & 998&-2.465&20&-1.698,662& 7&+1.864& 10\\
-2.820&5.313,951,84&1101&+2.351,42 & 998&-2.485&20&-1.698,655& 8&+1.874& 10\v
-2.825&5.313,940,83&1127&+2.361,40 & 997&-2.505&20&-1.698,647& 7&+1.884& 10\\
-2.830&5.313,929,56&1152&+2.371,37 & 998&-2.525&20&-1.698,640& 8&+1.894& 10\\
-2.835&5.313,918,04&1180&+2.381,35 & 998&-2.545&20&-1.698,632& 8&+1.904& 10\\
-2.840&5.313,906,24&1207&+2.391,33 & 997&-2.565&20&-1.698,624& 8&+1.914& 10\\
-2.845&5.313,894,17&1234&+2.401,30 & 998&-2.585&20&-1.698,616& 8&+1.924& 10\v
-2.850&5.313,881,83&    &+2.411,28 &    &-2.605&  &-1.698,608&  &+1.934&   \\
[\extrarowheight]\hline
\end{tabular}
\end{center}
\end{table*}
\begin{table*}
\begin{center}
\head, 4.
\small
\begin{tabular}{ | >{\quad$}c<{$\quad} |
>{\quad$}c<{$} >{$\hspace{1em}}r<{$\quad} |
>{\quad$}c<{$} >{$\hspace{1em}}r<{$\quad} |
>{\quad$}c<{$} >{$\hspace{1em}}r<{$\quad} |
>{\quad$}c<{$} >{$\hspace{1em}}r<{$\quad} |
>{\quad$}c<{$} >{$\hspace{1em}}r<{$\quad} |}
\hline\shim
\mathrm{Arg}
&\log\alpha&\hspace{\minus1em}\minus\Delta
&\log\beta&\hspace{\minus1em}\hphantom{\minus}\Delta
&\log\gamma\quad&\hspace{\minus1em}\hphantom{\minus}\Delta
&\log\alpha'&\hspace{\minus1em}\minus\Delta
&\log\beta'&\hspace{\minus1em}\hphantom{\minus}\Delta
\\[\extrarowheight]\hline\shim
-2.850&5.313,881,83&1264&+2.411,279&9974&-2.605&20&-1.698,608& 8&+1.934& 10\\
-2.855&5.313,869,19&1293&+2.421,253&9974&-2.625&20&-1.698,600& 9&+1.944& 10\\
-2.860&5.313,856,26&1323&+2.431,227&9974&-2.645&20&-1.698,591& 9&+1.954& 10\\
-2.865&5.313,843,03&1353&+2.441,201&9973&-2.665&20&-1.698,582& 9&+1.964& 10\\
-2.870&5.313,829,50&1385&+2.451,174&9972&-2.685&20&-1.698,573& 9&+1.974& 10\v
-2.875&5.313,815,65&1417&+2.461,146&9972&-2.705&20&-1.698,564&10&+1.984& 10\\
-2.880&5.313,801,48&1450&+2.471,118&9971&-2.725&20&-1.698,554& 9&+1.994& 10\\
-2.885&5.313,786,98&1484&+2.481,089&9970&-2.745&20&-1.698,545&10&+2.004& 10\\
-2.890&5.313,772,14&1518&+2.491,059&9970&-2.765&20&-1.698,535&10&+2.014&  9\\
-2.895&5.313,756,96&1553&+2.501,029&9969&-2.785&19&-1.698,525&11&+2.023& 10\v
-2.900&5.313,741,43&1590&+2.510,998&9968&-2.804&20&-1.698,514&10&+2.033& 10\\
-2.905&5.313,725,53&1626&+2.520,966&9968&-2.824&20&-1.698,504&11&+2.043& 10\\
-2.910&5.313,709,27&1664&+2.530,934&9966&-2.844&20&-1.698,493&11&+2.053& 10\\
-2.915&5.313,692,63&1702&+2.540,900&9966&-2.864&20&-1.698,482&11&+2.063& 10\\
-2.920&5.313,675,61&1742&+2.550,866&9965&-2.884&20&-1.698,471&12&+2.073& 10\v
-2.925&5.313,658,19&1783&+2.560,831&9965&-2.904&20&-1.698,459&12&+2.083& 10\\
-2.930&5.313,640,36&1824&+2.570,796&9963&-2.924&20&-1.698,447&12&+2.093& 10\\
-2.935&5.313,622,12&1866&+2.580,759&9963&-2.944&20&-1.698,435&12&+2.103& 10\\
-2.940&5.313,603,46&1909&+2.590,722&9962&-2.964&20&-1.698,423&13&+2.113& 10\\
-2.945&5.313,584,37&1953&+2.600,684&9961&-2.984&20&-1.698,410&13&+2.123& 10\v
-2.950&5.313,564,84&1999&+2.610,645&9960&-1.004&20&-1.698,397&13&+2.133& 10\\
-2.955&5.313,544,85&2045&+2.620,605&9959&-1.024&20&-1.698,384&14&+2.143& 10\\
-2.960&5.313,524,40&2093&+2.630,564&9958&-1.044&20&-1.698,370&14&+2.153& 10\\
-2.965&5.313,503,47&2141&+2.640,522&9957&-1.064&19&-1.698,356&14&+2.163& 10\\
-2.970&5.313,482,06&2191&+2.650,479&9956&-1.083&20&-1.698,342&15&+2.173& 10\v
-2.975&5.313,460,15&2241&+2.660,435&9956&-1.103&20&-1.698,327&15&+2.183& 10\\
-2.980&5.313,437,74&2293&+2.670,391&9954&-1.123&20&-1.698,312&15&+2.193& 10\\
-2.985&5.313,414,81&2347&+2.680,345&9953&-1.143&20&-1.698,297&16&+2.203&  9\\
-2.990&5.313,391,34&2400&+2.690,298&9952&-1.163&20&-1.698,281&15&+2.212& 10\\
-2.995&5.313,367,34&2457&+2.700,250&9951&-1.183&20&-1.698,266&17&+2.222& 10\v
-1.000&5.313,342,77&2513&+2.710,201&9950&-1.203&20&-1.698,249&17&+2.232& 10\\
-1.005&5.313,317,64&2571&+2.720,151&9948&-1.223&20&-1.698,232&17&+2.242& 10\\
-1.010&5.313,291,93&2631&+2.730,099&9948&-1.243&20&-1.698,215&17&+2.252& 10\\
-1.015&5.313,265,62&2691&+2.740,047&9946&-1.263&19&-1.698,198&18&+2.262& 10\\
-1.020&5.313,238,71&2754&+2.749,993&9945&-1.282&20&-1.698,180&18&+2.272& 10\v
-1.025&5.313,211,17&2818&+2.759,938&9943&-1.302&20&-1.698,162&19&+2.282& 10\\
-1.030&5.313,182,99&2883&+2.769,881&9943&-1.322&20&-1.698,143&19&+2.292& 10\\
-1.035&5.313,154,16&2949&+2.779,824&9941&-1.342&20&-1.698,124&20&+2.302& 10\\
-1.040&5.313,124,67&3018&+2.789,765&9939&-1.362&20&-1.698,104&20&+2.312& 10\\
-1.045&5.313,094,49&3087&+2.799,704&9939&-1.382&20&-1.698,084&20&+2.322& 10\v
-1.050&5.313,063,62&3159&+2.809,643&9936&-1.402&20&-1.698,064&21&+2.332& 10\\
-1.055&5.313,032,03&3232&+2.819,579&9936&-1.422&20&-1.698,043&22&+2.342&  9\\
-1.060&5.312,999,71&3306&+2.829,515&9934&-1.442&19&-1.698,021&22&+2.351& 10\\
-1.065&5.312,966,65&3383&+2.839,449&9932&-1.461&20&-1.697,999&22&+2.361& 10\\
-1.070&5.312,932,82&3460&+2.849,381&9931&-1.481&20&-1.697,977&23&+2.371& 10\v
-1.075&5.312,898,22&3541&+2.859,312&9929&-1.501&20&-1.697,954&24&+2.381& 10\\
-1.080&5.312,862,81&3623&+2.869,241&9928&-1.521&20&-1.697,930&24&+2.391& 10\\
-1.085&5.312,826,58&3706&+2.879,169&9926&-1.541&20&-1.697,906&25&+2.401& 10\\
-1.090&5.312,789,52&3791&+2.889,095&9924&-1.561&20&-1.697,881&25&+2.411& 10\\
-1.095&5.312,751,61&3879&+2.899,019&9922&-1.581&19&-1.697,856&26&+2.421& 10\v
-1.100&5.312,712,82&    &+2.908,941&    &-1.600&  &-1.697,830&  &+2.431&   \\
[\extrarowheight]\hline
\end{tabular}
\end{center}
\end{table*}

\end{document}